# Intertwined lattice deformation and magnetism in monovacancy graphene

Haricharan Padmanabhan[1,2,*] and B. R. K. Nanda[1]

[1]*Department of Physics, Indian Institute of Technology Madras, Chennai, India 600036* and
[2]*Department of Engineering Design, Indian Institute of Technology Madras, Chennai, India 600036*
(Dated: May 12, 2016)

Using density functional calculations we have investigated the local spin moment formation and lattice deformation in graphene when an isolated vacancy is created. We predict two competing equilibrium structures: a ground state planar configuration with a saturated local moment of 1.5 $\mu_B$, and a metastable non-planar configuration with a vanishing magnetic moment, at a modest energy expense of 50 meV. Though non-planarity relieves the lattice of vacancy-induced strain, the planar state is energetically favored due to maximally localized defect states (v$\sigma$, v$\pi$). In the planar configuration, charge transfer from itinerant (Dirac) states weakens the spin-polarization of v$\pi$ yielding a fractional moment, which is aligned parallel to the unpaired v$\sigma$ electron through Hund's coupling. As a byproduct, the Dirac states (d$\pi$) of the two sublattices undergo a minor spin-polarization and couple antiferromagnetically. In the non-planar configuration, the absence of orthogonal symmetry allows interaction between v$\sigma$ and local d$\pi$ states, to form a hybridized v$\sigma'$ state. The non-orthogonality also destabilizes the Hund's coupling, and an antiparallel alignment between v$\sigma$ and v$\pi$ lowers the energy. The gradual spin reversal of v$\pi$ with increasing non-planarity opens up the possibility of an intermediate structure with balanced v$\pi$ spin population. If such a structure is realized under external perturbations, diluted vacancy concentration may lead to v$\sigma$ based spin-1/2 paramagnetism. Carrier doping, electron or hole, does not alter the structural stability. However, the doping proportionately changes the occupancy of v$\pi$ state and hence the net magnetic moment.

## I. INTRODUCTION

Magnetism in graphene with defects in the form of vacancies has attracted a great deal of attention over the last decade, as the source of magnetism is the missing carbon atom rather than magnetically active elements such as transition metals [1–5]. The missing carbon atom leads to the formation of dangling $\sigma$ states and a quasilocalized (zero-mode [6]) $\pi$ state which together form local moments to induce magnetism in the system.

A review of theoretical (primarily ab-initio) literature reveals that vacancy-induced lattice distortion and magnetism are closely related to each other. A large fraction of reports [2–4, 7] support the formation of a ground state with a planar lattice structure, while some [2, 8–11] have identified a metastable non-planar configuration in addition to this. When the lattice distortion is restricted to the graphene plane, magnetic moments in the range of 1-2 $\mu_B$ have been predicted, whereas in the non-planar configuration, most reports [1, 2, 8, 11, 12] suggest a non-magnetic state. However, the latter has been contested by a few [9, 10], who argue that local moments do exist in the non-planar configuration, and are aligned antiparallel to each other resulting in a reduced net magnetic moment. Experimental reports [5, 13, 14] have confirmed the existence of local magnetic moments, and the magnetism in the system has been characterized [14] as v$\sigma$-based spin-1/2 paramagnetism.

Despite being largely identified as a metastable state within density functional studies, with a few exceptions [1, 10], the non-planar configuration of the vacancy defect assumes significance in the presence of external factors that stabilize it as the ground state. This includes rippling at finite temperatures, doping [9], isotropic compression [15, 16], interaction with substrates [11], binding to foreign species [9, 17], and interaction with experimental probes [1]. Some of these reports [11, 16] also indicate that the degree of out-of-plane distortion around the vacancy can be controlled by these external factors.

In this article we have performed electronic structure calculations with a focus on the non-planar configuration in order to address the conflicting results in literature, as discussed above. In particular, we examine whether formation of local moments is favored in the non-planar configuration, and if so, explain why reduced magnetic moments are predicted. We revisit the mechanism of spin-polarization in the planar configuration and study the role of the vacancy as an electron acceptor in order to explain the reason behind varied magnetic moments reported in literature. Finally, as various external factors may influence the nature and extent of lattice deformation, we consider it necessary to study the development of the electronic and magnetic structure as the system evolves from the planar to the non-planar configuration.

In the planar configuration, we find that the vacancy-induced $\sigma$ states (v$\sigma$) contribute 1 $\mu_B$. On the other hand, the quasilocalized $\pi$ state (v$\pi$) has only a fractional contribution to the magnetic moment as its occupancy exceeds one due to charge transfer from the Dirac states (d$\pi$). The charge transfer also creates itinerant spins that are antiferromagnetically coupled between the two sublattices, a feature that has not been identified so far. The v$\pi$ and v$\sigma$ local moments are aligned parallel to each other due to Hund's coupling. With out-of-plane lattice deformation around the vacancy, three processes occur simultaneously to drastically alter the electronic structure of the system. Firstly, the orthogonality be-tween the σ and π states is broken, and a covalent interaction between vσ and local dπ is initiated, producing a hybridized vσ′ state. Secondly, the charge transfer is reduced, resulting in an unpaired vπ electron. Finally, the non-planarity destabilizes the Hund's coupling between the local moments, driving them towards an antiparallel alignment and leading to a vanishing magnetic moment in the metastable non-planar configuration. Due to the gradual spin-reversal of vπ, an intermediate structure may be realized in which the state is equally occupied in both spin channels. With a dilute vacancy concentration, this provides a window for σ based spin-1/2 paramagnetism.

*Computational Details*: Density functional calculations are performed using Vanderbilt ultra-soft pseudopotentials [18] and plane wave basis sets as implemented in Quantum Espresso [19]. Exchange-correlation potential is approximated through the PBE-GGA [20] functional. The kinetic energy cutoff to fix the number of plane waves is taken as 30 Ry, and the kinetic energy cutoff for charge density is taken as 250 Ry. Full structural relaxation was carried out for the relevant structures, with a low convergence threshold of $10^{-6}$ Ry per atom. In order to accurately describe the electronic structure of defect-induced states, a dense 24x24x1 k-mesh yielding 86 irreducible k-points in the Brillouin Zone for the unit cell is used. This is found to estimate total energy and magnetic moments with good accuracy within the pseudopotential approximation. Finally, a small Gaussian smearing of 0.001 Ry was used to enable speedy convergence of self-consistent calculations.

All calculations were performed using supercells consisting of 6x6 unit cells of graphene, with a single vacancy. This corresponds to an intervacancy distance of 14.8 Å, allowing for sufficient isolation to study vacancy-induced lattice distortions and magnetism.

## II. RESULTS AND DISCUSSION

This section is organized into three parts. First, the electronic structure of the planar configuration of graphene with a monovacancy will be reexamined, to provide a base to analyse and compare subsequent results with. The second part will feature a detailed analysis of the electronic structure and magnetism of the non-planar configuration. Finally, the energetics affecting the out-of-plane displacement of the atoms in the vicinity of the vacancy will be discussed.

### A. Electronic structure of planar configuration

A complete structural relaxation of the system results in a ground state possessing a planar lattice structure, as shown in Figure 1(a). The vacancy results in three dangling σ bonds, which break their degeneracy by initiating covalent interactions and inducing a planar Jahn-Teller distortion [4, 21]. This distortion brings the C2-C3 atoms closer to each other, while the apical carbon atom (see Figure 1(c)) moves away from it, forming an isosceles triangle around the vacancy. The lattice structure in the local neighborhood is minimally distorted and does not affect the electronic structure significantly [3, 4, 7].

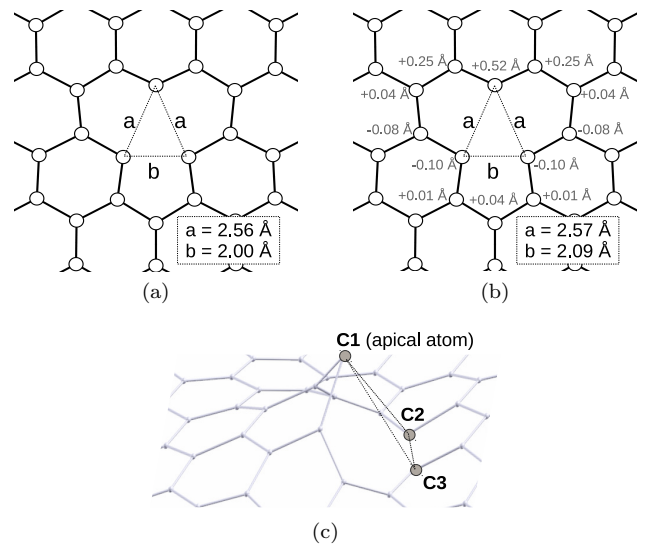

FIG. 1. Equilibrium lattice structures for (a) the ground state (planar) and (b) the metastable state (non-planar). The displacements of the atoms along the z-direction are indicated in the figure. The side view of the non-planar structure showing local rippling induced by the vacancy is shown in (c). Here, out-of-plane displacements are exaggerated by a factor of 5 for clarity.

The salient features of the electronic structure, shown in Figure 2, include a vacancy-induced dangling σ state (vσ) on the apical carbon atom, which generates a local spin moment. In addition to this, there is a quasilocalized vacancy-induced π state (vπ), resulting from the bipartite nature of the graphene lattice [6]. The vπ state, also known as the zero-mode state, while localized in k-space, is spread over the majority sublattice in real space, decreasing in intensity ($\propto \frac{1}{r^2}$) away from the vacancy. This state undergoes partial spin-polarization, giving rise to a fractional magnetic moment. Since the system is planar, the vσ and the π states are orthogonal to each other, resulting in a parallel alignment of their spins as favored by Hund's coupling.

The process of partial spin-polarization of the vπ state needs to be understood in detail, as it has a significant impact on the formation of local moments. In Figure 2(a), the band structure is shown in the vicinity of the Fermi level. The figure shows that the vacancy breaks the electron-hole symmetry of the Dirac band (dπ) by pushing it above the Fermi level near the Dirac point. This results in an internal charge transfer from dπ to the weakly dispersive vπ state. As a consequence, the latter is now allowed to occupy more than one electron, which weakens its spin-polarization. We note that if vπ had been an ideal, dispersionless zero-mode state, such a charge transfer would not have been possible because of the binary nature of its occupancy.

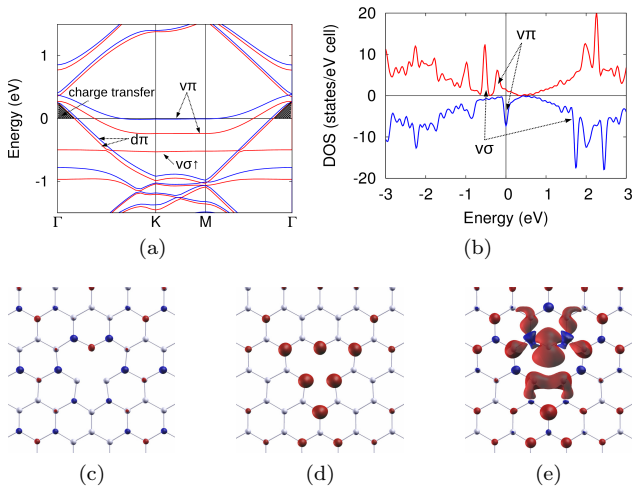

FIG. 2. (Color online) Red (blue) represents spin-up (spin-down) channel here as well as in other relevant figures. (a) Band structure near the Fermi level showing the dangling vσ, quasilocalized vπ and the itinerant Dirac (dπ) states of the ground state planar configuration. The shaded portions indicate the amount of charge transfer from the dπ to vπ states. (b) The spin-polarized density of states. (c) The itinerant spin density showing the antiferromagnetic coupling between the two sublattices. (d) The π spin density (vπ+dπ) and the total spin density (vπ + dπ + vσ) are shown in (d) and (e) respectively. The isovalues for the spin density plots have not been mentioned since different, but appropriate energy ranges have been chosen to isolate charge densities of different states.

The charge transfer also spin-polarizes the otherwise non-magnetic conducting states. The spin-density, plotted in Figure 2(c), shows that the itinerant spins exhibit a weak antiferromagnetic coupling between the two sublattices. This, superimposed on the dominating density of the aligned local spins of the vπ and vσ states, results in the overall ferrimagnetic spin-density pattern as seen in Figure 2(d) and Figure 2(e). This has been mistakenly interpreted in some reports [4] as a Kondo-like antiferromagnetic coupling between the local spins and the conduction band.

The total magnetic moment induced by the vacancy is thus $(1 + x)\ \mu_B$, where vσ contributes 1, and x is the contribution due to vπ. As explained in the previous passage, the fractional contribution decreases with increase in charge transfer from the conduction band, a process that cannot be exactly quantified within band theory. This, along with the supercell effect discussed by others [22] is the reason for the wide range of magnetic moments predicted in literature [3, 4, 7]. For the 6x6 supercell used in this work, a total magnetic moment of 1.5 $\mu_B$ was obtained. Our fixed spin moment calculations (shown in Figure 4), which will be discussed in detail in the next section, reveals a shallow minimum at 1.5 $\mu_B$, which indicates that while the system has a magnetic ground state, the magnetic moment is ill-defined even at low temperatures.

### B. Non-planar vacancy defect

While the ground state of monovacancy graphene is found to be planar, a metastable state that features significant out-of-plane displacements is observed. This metastable state lies only 54 meV above the ground state, and therefore assumes significance due to the possibility of stabilization by external factors [9, 11, 15, 16]. One of the characteristic features of the lattice structure of this state, shown in Figure 7, is a large out-of-plane displacement of the apical carbon atom (labeled C1 in Figure 1(c)). When the structure is completely relaxed, C1 is found to be displaced above the plane by 0.52 Å. The surrounding atoms are displaced below the plane by 0.1 − 0.2 Å, resulting in a local rippling as shown in Figure 1(c). The other structural features observed in the relaxed planar configuration are largely preserved. The driving force behind the out-of-plane displacement is the negative strain induced by the vacancy defect, an aspect that will be discussed in more detail in the next section.

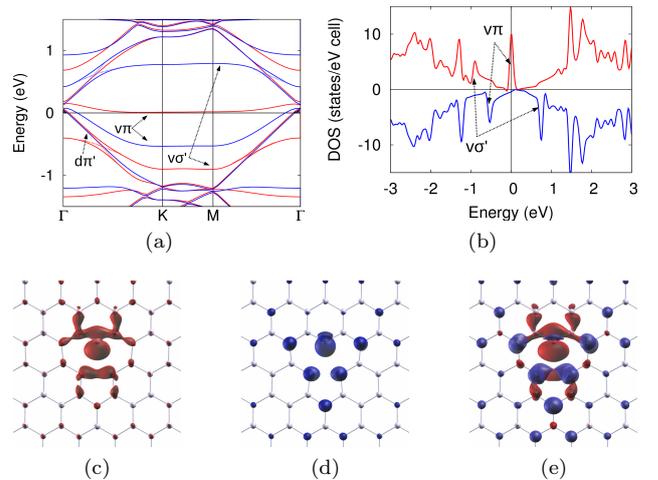

FIG. 3. (Color online) (a) The band structure near the Fermi level of the metastable non-planar configuration. Hybridization between vσ and dπ leads to band-splitting around Γ. The dotted lines at the band splitting are artificially constructed to act as a guide for the eyes to distinguish the hybridized vσ′ and dπ′ states. (b) The spin-polarized density of states showing the antiparallel coupling between vπ and vσ′. The vσ′ spin density, vπ spin density, and total spin density (vσ′ + vπ + dπ′) are shown in (c), (d), and (e) respectively. The isovalues for the spin density plots have not been mentioned since different, but appropriate energy ranges have been chosen to isolate charge densities of different states.

The non-planar configuration is found to form vacancy-induced π and σ unpaired spins following a mechanism similar to that in the ground state planar configuration. However, the electronic and magnetic structure is drastically altered by the out-of-plane deformation, as can be seen in Figure 3. We first observe that the v$\sigma$ state is no longer purely $\sigma$ in character. The out-of-plane displacement breaks the orthogonality between the $\sigma$ and $\pi$ orbitals, allowing them to interact covalently. The original v$\sigma$ interacts with d$\pi$, resulting in a more dispersed, hybridized state which we label as v$\sigma'$. Like v$\sigma$, this state remains singly occupied. The electron-hole symmetry of the conduction band is largely retained in the non-planar configuration, unlike in the ground state, and therefore the charge transfer from the Dirac band to v$\pi$ is absent or negligible. This results in v$\pi$ having an occupation one, leading to almost complete spin-polarization, and a corresponding unpaired spin.

Apart from v$\sigma$-d$\pi$ hybridization, there is another minor change in the conduction band. A comparison of band structures in Figure 2(a) and Figure 3(a) shows that the spin minority channel is slightly more occupied than the spin majority channel in the non-planar configuration. This resembles a weak antiferromagnetic coupling between d$\pi$ and v$\sigma$, which has been identified by some [10] as Kondo coupling.

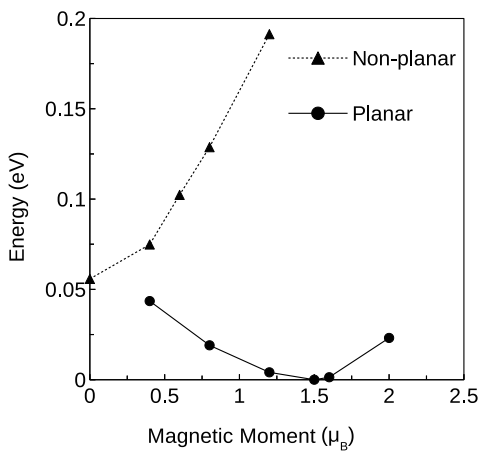

FIG. 4. The relative total energy plotted as a function of the magnetic moment for the planar and non-planar equilibrium configurations. The results were obtained using fixed spin moment calculations.

Since v$\pi$ and v$\sigma$' are no more orthogonal, Hund's coupling is not favored, and instead, an antiparallel alignment of these spins occurs, so as to minimize kinetic energy. The spin alignment can be seen in the density of states and spin-density plots shown in Figure 3. Consequently, the magnetic moment corresponding to this system almost vanishes (0.06 $\mu_B$). The stability of the antiparallel coupling is quantified in Figure 4, which is obtained using the fixed spin moment (FSM) method. We observe that a Hund's coupling would cost 0.3 eV over an antiparallel alignment.

We would like to emphasize that contrary to many reports [1, 2, 8, 11, 12, 16], the ground state of the non-planar configuration is not non-magnetic. Our calculations show that the non-magnetic state in fact lies 55 meV higher than the state with antiparallel spin structure discussed above.

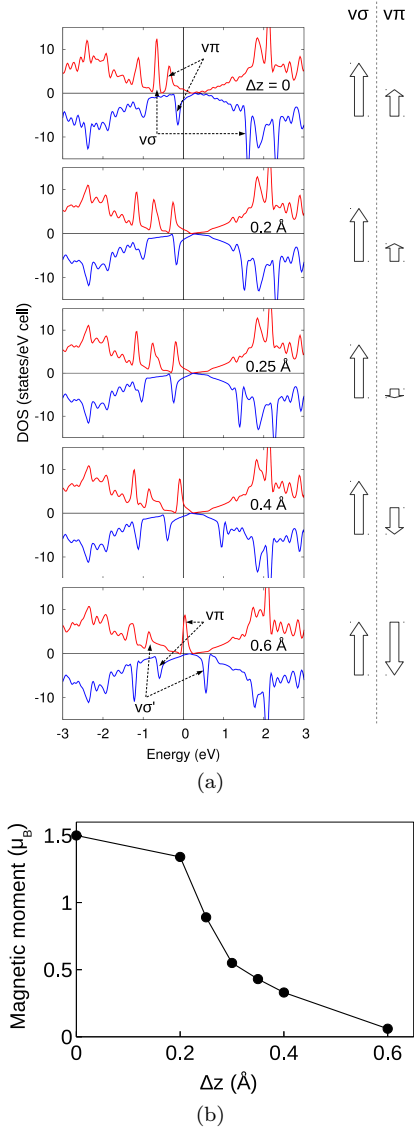

FIG. 5. (Color online) (a) Density of States plots showing the evolution of the electronic structure as the system moves from the ground state planar lattice to the metastable state non-planar lattice. The results are obtained by carrying out a virtual experiment where the apical carbon atom (C1) is gradually displaced in the out-of-plane (z) direction from the planar case of 0 Å to a maximum of 0.6 Å. The magnitude of displacement ($\Delta z$) is indicated in the respective figures. The gradual change in the spin alignment of v$\sigma'$ and v$\pi$ moments from parallel to antiparallel is schematically shown on the right. (b) The net magnetic moment with respect to $\Delta z$.

In the previous two parts we have discussed the ground state planar configuration and the metastable non-planar configuration. Since it is found that the energy difference between these two states is reasonably small (∼ 55 meV), a practical structure may lie somewhere in between the two under different external factors such as compression [16] and substrate interaction [11]. Therefore it is necessary to study the gradual change of the electronic and magnetic structure as the system is taken from the planar to the metastable non-planar configuration. In this context, we have carried out a virtual experiment where the apical carbon atom (C1, see Figure 1(c)) is gradually displaced along the out-of-plane direction from the planar case to a maximum of 0.6 Å. Such an experiment is acceptable, since it reproduces all the salient features of the electronic structure of the concerned systems with reasonable accuracy including the spin-polarized vacancy-induced states.

To explain the evolution of the electronic and magnetic structure, we have plotted the spin-polarized DOS for different out-of-plane displacements of C1 in Figure 5(a). The following observations are made from the figure: firstly, the v$\sigma$ state becomes increasingly delocalized with out-of-plane displacement due to covalent interaction with the conduction band. Note that it still represents a single unpaired spin. Secondly, the charge transfer from the conduction band gradually reduces, so that the occupancy of v$\pi$ reduces to one for the metastable case. Finally and most importantly, the out-of-plane displacement changes the spin-polarization of the original v$\pi$. It is gradually pushed below in energy in the spin-down channel and pushed above in the spin-up channel, as seen in Figure 5(a). For a small range of intermediate out-of-plane displacements, the v$\pi$ state is found to have almost equal occupancy in both the spin channels, making it magnetically inactive. This leaves the system with a single v$\sigma$ local moment. With a low vacancy concentration, this may result in spin-1/2 paramagnetism, which coincides with some experimental observations [14]. To give a quantitative perspective, we have plotted the net magnetic moment with respect to out-of-plane displacement in Figure 5(b). Our DFT calculations suggest that spin-1/2 paramagnetism would be obtained if the out-of-plane displacement lies between 0.2 and 0.3 Å.

With the maximum out-of-plane displacement of 0.6 Å, the spin-polarization of the v$\pi$ state is completely reversed, and it is aligned antiparallel to the v$\sigma'$ state. Since both the states are singly occupied, this results in a vanishing magnetic moment.

### C. Carrier doping and spin-polarization

In the last section, we showed that both v$\sigma$ and v$\pi$ states are crucial to both spin-polarization and structural deformation. Therefore, it is assumed that the varying occupancy of these two states can affect both the lattice structure and magnetism. In this context, here we have studied the electronic structure and structural stability when the system is doped with one hole or one electron. From the total energy calculations, we find that the ground state still favors a planar arrangement of the atoms in the neighborhood of the vacancy. Hole and electron doping make the defined non-planar structure unstable by 46 meV and 210 meV respectively, as compared to the planar structure.

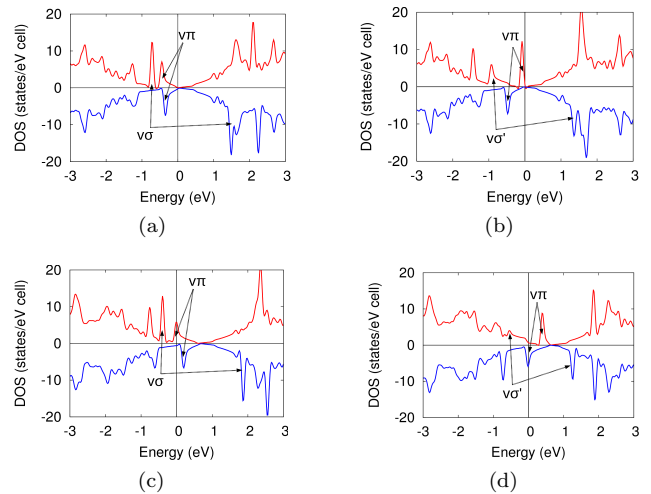

FIG. 6. (Color online) The top and bottom panels show the spin-polarized DOS for one electron and one hole doped systems respectively. The left (a and c) represent the planar ground state, and the right (b and d) represent the non-planar metastable state. In the case of electron doping, the v$\pi$ state is completely occupied to give 1 $\mu_B$. In the case of hole doping, the v$\pi$ state is partially occupied in one spin channel only. Other features of the electronic structure shown in Figure 2 and Figure 3 are largely retained.

To analyze the effect of carrier doping on the electronic structure, in Figure 6 we have plotted the carrier-doped spin-polrized DOS both for the planar and non-planar configurations. The figure shows that the features of the electronic structure remain largely unchanged from that of the undoped system. Doping primarily changes the occupancy and hence spin-polarization of the v$\pi$ state. Doping of one electron increases its occupancy from one to two. Thus the v$\pi$ state is completely occupied (see Figure 6(a) and 6(b)) and only the v$\sigma$ state has one unpaired spin, as in the undoped system. Similarly doping of one hole decreases the occupancy from 1 to $\delta$. Now, since v$\pi$ and v$\sigma$ states are parallel in the planar ground state, the net magnetic moment of the system is 1+$\delta$. In the metastable non-planar configuration, the antiparallel alignment of the v$\sigma$ and v$\pi$ spins yields a magnetic moment of 1-$\delta$ to the system. Ideally $\delta$ is expected to be zero. However, we find it to be 0.36.

### D. Strong correlation effect on the electronic structure

The narrow band width of the vacancy induced states leads to localization. In transition metal oxides(TMOs)

TABLE I. Energetics and local moment formation with and without U, both for planar and metastable non-planar configuration. The total energy values of the planar and non-planar structures are expressed in relative terms with respect to a given U.

|  | Planar | | Non-planar | |
|---|---|---|---|---|
| U (eV) | Energy (meV) | MM[a] ($\mu_B$) | Energy (meV) | MM[a] ($\mu_B$) |
| 0 | 0 | 1.5 | 55 | 0.06 |
| 3 | 0 | 2.0 | 52 | 0.4 |

[a] Local magnetic moment

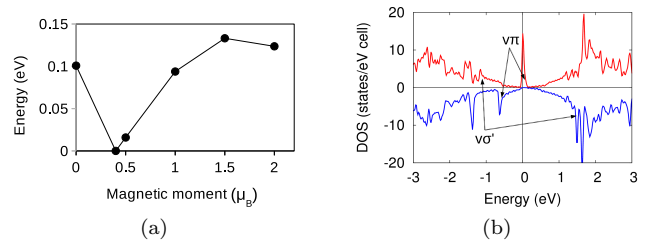

FIG. 7. (a) Total energy with respect to fixed spin moment for the metastable state, with U = 3 eV. The minimum energy state is achieved at 0.4 $\mu_B$. The DOS corresponding to the minimum energy state is shown on the right.

such as perovskite manganites [23], the localized d-states are strongly correlated. In fact, the antiferromagnetic and insulating behavior of these compounds are primarily attributed to the strong correlation effect. One way to predict the ground state in strongly correlated systems is to carry out DFT+U calculations. For the TMOs, U = 3 eV is found to be good enough to obtain the ground state[24].

To find out if the localized vacancy states in monovacancy graphene are strongly correlated, and whether they affect the process of spin-polarization, we have calculated the DFT+U (U = 3 eV) electronic structure. In Table I, a comparison is made between the energetics and local moments obtained through DFT and DFT+U for both planar and non-planar configurations.

From the table we find that the ground state still favors the planar configuration. With the inclusion of U, the magnetic moment is increased as expected. In the case of planar configuration, the spin-polarization is complete and the system yields 2 $\mu_B$. To determine the minimum energy state for the meta-stable non-planar configuration, we have plotted the total energy as a function of magnetic moment in Figure 7(a). The results are obtained from fixed-spin-moment (FSM) calculations. The plot shows that the energy is minimum when the net magnetic moment is close to 0.4 $\mu_B$, which is 0.34 $\mu_B$ higher than that of the DFT-only results (see Table I). This increase in magnetic moment is attributed to the increase in occupancy of the v$\pi$ state as can be seen from the spin-polarized DOS plotted in Figure 7(b). Without U, we have predicted in subsection (A) that there is a certain charge transfer from the v$\pi$ to d$\pi$ state (see Figure 3(a) and (b)). With inclusion of U the charge transfer reduces, resulting in an increase in the v$\pi$ occupancy. However, the v$\pi$ and v$\sigma$ spins still remain antiparallel for the non-planar configuration, as already discussed in subsection-B

### E. Energetics of out-of-plane displacement

So far, we have seen that minor deformation in the lattice, which is expected during synthesis under different external conditions, results in large changes in the electronic structure. In this regard, a study of the process of stabilization as the system makes a transition from the ground state planar to the metastable non-planar configuration will provide a useful insight to how the electronic structure may be experimentally manipulated by external factors. In general, the stability of the system is determined by two entities - ions and valence electrons. In this context we have considered the Madelung energy as a measure of the ion-ion interaction, and the rest of the energy terms provided by DFT are attributed to the electrons.

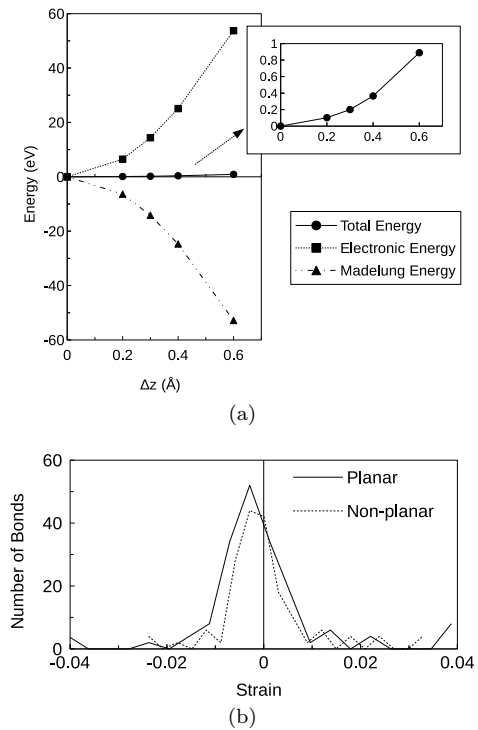

FIG. 8. The various energy contributions with respect to out-of-plane deformation using the model discussed in the text is shown in (a), with the total energy magnified in the inset for clarity. (b) Histogram of bond lengths in the planar and non-planar equilibrium configurations.

The virtual experiment, described in the previous section, i.e. by mapping the degree of non-planarity to the out-of-plane displacement of the apical atom (C1) only, will help us in analyzing the ionic and electronic contributions to the stability of the system. The C1 atom is gradually displaced along the out-of-plane direction from the planar case to a maximum of 0.6 Å. The estimated Madelung energy, electronic energy, and the total energy as a function of out-of-plane displacement are shown in Figure 8(a). We observe that the Madelung energy decreases with the out-of-plane displacement, which suggests that it is favorable for the lattice to stabilize in a non-planar configuration. From this, we also infer that the non-planar distortion relieves the lattice of negative strain energy induced by the vacancy, consistent with observations made by [1, 25, 26]. A histogram of the C-C bond lengths in the ground state planar and metastable non-planar configurations, shown in Figure 8(b), supports this claim. According to the figure, a larger fraction of bonds are found to be close to the pristine C-C bond length ($a_0 = 1.424$ Å) in the non-planar configuration than in the planar configuration. The mean bond length in the planar configuration is lower than $a_0$, reflecting the negative strain induced by the vacancy in the lattice.

In contrast to the ionic energy, the contribution due to the electrons increases with out-of-plane displacement (Figure 8(a)). The magnitude of stabilization of the planar configuration by the valence electrons is large enough to offset the energy penalty due to its higher lattice strain. As a consequence, the lattice deformation is restricted within the plane, resulting in a planar ground state.

Note that though the total energy monotonically increases with out-of-plane displacement in this virtual experiment (Figure 8(a) inset), we would in reality expect a local minimum corresponding to the metastable non-planar configuration, at around 0.5 Å. This feature has not been reproduced, as the out-of-plane displacement was restricted to only C1, with the local rippling being neglected.

### III. CONCLUSION

To summarize, we have explained the role of lattice deformation in charge reconstruction and in turn, spin polarization of defect states in monovacancy graphene. We show that irrespective of the degree of non-planarity of the lattice, the dangling $\sigma$ states around the vacancy collectively contribute one Bohr Magneton to the system. On the other hand, the spin-polarization of the quasilocalized $\pi$ state varies with the extent of out-of-plane distortion in the immediate neighborhood of the vacancy. In the ground state planar structure v$\pi$ is fractionally polarized and aligned with the orthogonal v$\sigma$ spin. Increasing the out-of-plane distortion gradually reverses the v$\pi$ spin occupancy, resulting in a vanishing magnetic moment in the metastable non-planar structure. The (Kondo-like) antiferromagnetic coupling between the local and itinerant spins is found to be negligible or absent. Electron and hole doping, intentional or otherwise, doesn't affect the lattice stability, but proportionately changes the occupation of v$\pi$ and hence the net local moment. We also find that inclusion of Hubbard U neither affects the lattice nor the coupling between v$\pi$ and v$\sigma$ spins. However, it enhances the magnetization, as expected. Different experimental works have predicted spin-1 [27] or spin-1/2 [14] paramagnetism in vacancy graphene. Our calculations convincingly show that spin-1 paramagnetism is realized when the planar symmetry of the graphene lattice is preserved, and a certain degree of non-planarity introduces spin-1/2 paramagnetism in the system.

### ACKNOWLEDGEMENTS

The work is funded by the IIT Madras exploratory research program. The authors acknowledge the computational resources provided by HPCE IIT Madras. H.P. thanks Ajit Jena and Sajid Ali for help with computation.

*Current address - Department of Materials Science and Engineering, Pennsylvania State University, University Park, Pennsylvania 16802